# Direct-drive ocean wave-powered batch reverse osmosis

Katie M. Brodersen[a], Emily A. Bywater[a], Alec M. Lanter[a], Hayden H. Schennum[a], Kumansh N. Furia[a], Maulee K. Sheth[a], Nathaniel S. Kiefer[a], Brittany K. Cafferty[a], Akshay K. Rao[a], Jose M. Garcia[b], David M. Warsinger[a]*

*Corresponding author

[a]School of Mechanical Engineering and Birck Nanotechnology Center, Purdue University, West Lafayette, IN, 47907, USA.

[b]School of Engineering Technology, Purdue University, West Lafayette, IN, 47907, USA.

## Abstract

Ocean waves provide a consistent, reliable source of clean energy making them a viable energy source for desalination. Ocean wave energy is useful to coastal communities, especially island nations. However, large capital costs render current wave-powered desalination technologies economically infeasible. This work presents a high efficiency configuration for ocean wave energy powering batch reverse osmosis. The proposed system uses seawater as the working fluid in a hydro-mechanical wave energy converter and replaces the reverse osmosis high-pressure pump with a hydraulic converter for direct-drive coupling. This allows for minimal intermediary power conversions, fewer components, and higher efficiencies. The concept was analyzed with MATLAB to model the transient energy dynamics of the wave energy converter, power take-off system, and desalination load. The fully hydro-mechanical coupling, incorporating energy recovery, could achieve an SEC and LCOW as low as 2.30 kWh/m³ and $1.96, respectively, for different sea states. The results were validated at the sub-system level against existing literature on wave energy models and previous work completed on batch reverse osmosis models, as this system was the first to combine these two technologies. SEC and LCOW values were validated by comparing to known and predicted values for various types of RO systems.

## Graphical Abstract

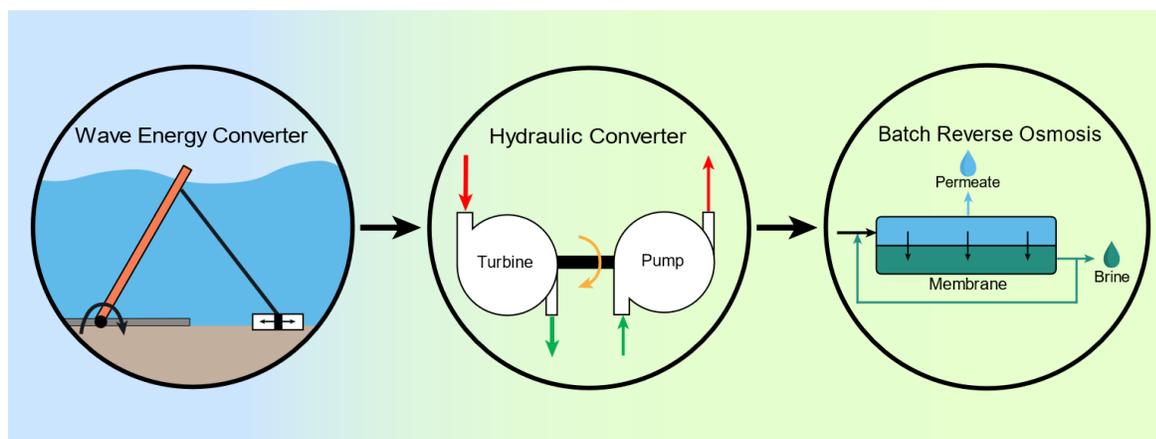

## Highlights

- ☐ The first direct-drive wave-energy powered batch reverse osmosis configuration.
- ☐ Wave-powered batch reverse osmosis (WPBRO) shows high energy efficiency.
- ☐ Robust dynamic models show WPBRO performance with various sea states.
- ☐ System models explore design considerations for scaling a WPBRO system.
- ☐ Specific energy consumption and levelized cost of water are shown to be competitive.

## Nomenclature

**RO:** Reverse osmosis

**BRO:** Batch reverse osmosis

**CRO:** Continuous reverse osmosis

**AWP:** Annual water production

**CapEx:** Capital expenditures

**OpEx:** Operational expenditures

**WPBRO:** Wave-powered batch reverse osmosis

**RES:** Renewable energy sources

**RR:** Recovery ratio

**SEC:** Specific energy consumption

**PV-RO:** Photovoltaic reverse osmosis

**WEC:** Wave energy converter

**OSWEC:** Oscillating surge wave energy converter

**PTO:** Power take-off

**FCD**: Flow control device

**PD:** Proportional derivative

**LCOW:** Levelized cost of water

## **Introduction**

While two-thirds of the earth are covered by water [1], only 1% of surface water is suitable for domestic and industrial purposes, and far less can be used sustainably [2]. Presently, more than a quarter of the world's population lacks access to sufficient purification facilities [3], which will only be exacerbated with population growth, climate change, and increased agricultural needs [4]. According to the United Nations World Water Development Report (2021) [5], over 40% of people will face water scarcity by 2030. As Africa in particular faces surface and groundwater depletion, the 35 African countries bordering a seafront may look to desalination as a solution [6]; however, progress in this direction has been obstructed by a lack of financial and energy resources to power traditional desalination systems, which are not widely available in most of the continent [7].

The desalination market has grown in capacity by 20% between 2016 and 2020 [8], and it will continue to grow as population increases and freshwater sources are depleted. However, rising interest in desalination has drawn attention to concerns about its high energy requirements. With the detrimental impact of fossil fuels on the environment, clean renewable energy sources (RES) are desirable alternatives for powering desalination systems. In addition to energy use's environmental effects, energy-intensity is also a financial burden. Energy consumption makes up the largest section of operational expenditures for water desalination, at approximately 36% of total operational expenditures for a typical plant [8]. Off-grid communities reliant on diesel generators to drive their desalination plants could pay anywhere from $3.00 to $8.00/m$^3$ for fresh water [9]. There is a significant need for renewable-driven desalination [10].

*Batch Reverse Osmosis Desalination*

The most common desalination process is reverse osmosis (RO), which accounts for 69% of the volume of water desalinated [11]. In traditional continuous RO (CRO), seawater traverses multiple RO membrane stages at a constant high-pressure and brine is discharged at the end of the process. The specific energy consumption (SEC) to drive a CRO process with standard conditions, seawater with 35 g/kg salinity and 50% recovery ratio (RR), ranges from 1.91 kWh/m$^3$ to 4.0 kWh/m$^3$ depending on the capacity of the plant [12]. Innovations may allow the SEC to approach closer to the thermodynamic limits for these conditions, 1.09 kWh/m$^3$ [13]. In contrast to CRO, batch desalination processes like batch reverse osmosis (BRO) and closed-circuit reverse osmosis (CCRO) recirculate the brine while varying the applied pressure along with the osmotic pressure. These processes greatly reduce the energy requirement as compared to CRO and provide additional benefits like biocidal salinity cycling and high recovery capabilities [12, 14-19]. Prior work has considered practical methods of achieving BRO with conventional components like pressure exchangers and piston-cylinders, which have been modeled to achieve an SEC of 1.88 kWh/m$^3$, even at low capacities [12]. BRO has also been shown to be staged and operated as an osmotically assisted process, called batch counter-flow reverse osmosis (BCFRO), to handle higher salinities and recoveries [15]. Therefore, there is merit in considering how BRO may be integrated in new configurations to make additional gains in efficiency.

*Renewable-Driven Desalination*

Several methods of driving RO with RES have been studied [20-22]. Photovoltaic (PV) solar desalination with battery energy storage is dominant in RES-powered desalination due to its cost-effectiveness and flexibility for large and small systems [23, 24]. However, PV-RO is constrained to its periodic and relatively low availability as well as the large land footprint required for PV to generate adequate energy. Wind energy is relatively inexpensive and has low environmental impact but is limited by a substantial land footprint and intermittent availability, much like solar energy. Wind energy is second only to solar energy in its use as a RES to power desalination [2]. Additionally, geothermal energy is highly stable and reliable, as it produces a consistent heat flux. It has low operational costs due to its independence of atmospheric and temporal patterns but is limited by its minimal availability and the high capital expenses of geothermal power plants [25]. Recent efforts have been aimed to incorporate salinity gradient energy storage and energy production in dynamic reverse osmosis processes [17, 26, 27]. These systems have reliable, long-term energy storage but have some concerns regarding economic feasibility.

A readily available RES for seawater desalination is marine energy because of its proximity to the intake of seawater reverse osmosis (SWRO) systems. Marine energy comprises ocean thermal energy and mechanical energy from waves and currents [8]. It is more stable than solar and wind energy because of its high energy density and consistency [28]. Marine energy also provides the opportunity for direct hydraulic power take-off, or conversion of energy from water to work-consuming and work-producing devices. This increases system efficiency by eliminating several energy conversion steps and reducing the cost of materials [8]. Additionally, the land footprint used by this RES is negligible. While promising, marine energy technologies have not yet been commercialized on a large scale [29]. Their market value is not yet competitive with solar and wind energy, as the levelized cost of water (LCOW) for wave-powered RO is higher than PV-RO and wind-powered RO. However, the market need is present. Remote island and coastal communities are often reliant on the high cost of imported diesel fuel and/or water to meet their needs. Power instability is another risk in remote regions, where less-resilient grids are vulnerable to interruptions during storms [8]. Consistently available and reliable marine energy mitigates these risks.

*Wave-Powered Desalination*

At present, 40% of the world population lives within 100 kilometers of a coastline [30]. Harnessing the energy-dense and locally available resource of ocean waves to power seawater RO is a sensible solution for coastal water scarcity [31, 32]. When selecting a wave-powered RO system, the mechanical and cost efficiency of different wave-powered desalination systems can be used to evaluate their performances. A leading wave-powered desalination company, Resolute Marine, estimates an LCOW of $1.30/m³ for their Wave2O™ system which uses a surge converter WEC on the seafloor to pressurize water to drive RO onshore [33]. A pressure-exchanger energy recovery device is used to reduce the energy consumption of RO [9]. Another competitor, Wavepiston, uses a chain of moving plates near the surface to pump seawater through a pipe to an onshore RO system, for an estimated LCOW of $1.80/m³ [34]. In 2017, NREL researchers conducted a baseline study of WEC desalination farms and arrived at $1.82/m³ for a system that generates 3100 m³/day of water. The specific energy consumption (SEC) for this study was estimated as 2.8 kWh/m³ [35].

Figure 1 illustrates the key differences between each system. All three systems use surge converter WECs to drive RO, but the power take-offs of each WEC are configured differently. The surge WECs in Resolute Marine and WPBRO are very similar and harness energy from linear and rotational motion at the seafloor, while the WEC used by Wavepiston captures energy from linear motion near the surface. Wavepiston directly sends the seawater pressurized by the linear motion to an accumulator, ERD, and finally RO through an adaptive hydraulic pressure developer integrated with the WEC [36]. Like most WEC-RO systems, Resolute Marine and Wavepiston use CRO with an energy recovery device (ERD) for desalination. WPBRO is the first direct integration of a WEC with BRO. For the couplings, Resolute Marine and WPBRO use the pressurized water from the WEC to drive a turbine to treat seawater drawn from a beach well, dispensing the WEC water back to the sea. The additional advantage of the WPBRO coupling is that flow control devices (FCDs) are implemented to improve dampening of the nonlinear energy profile from the WEC. The full WPBRO system is shown in Figure 2.

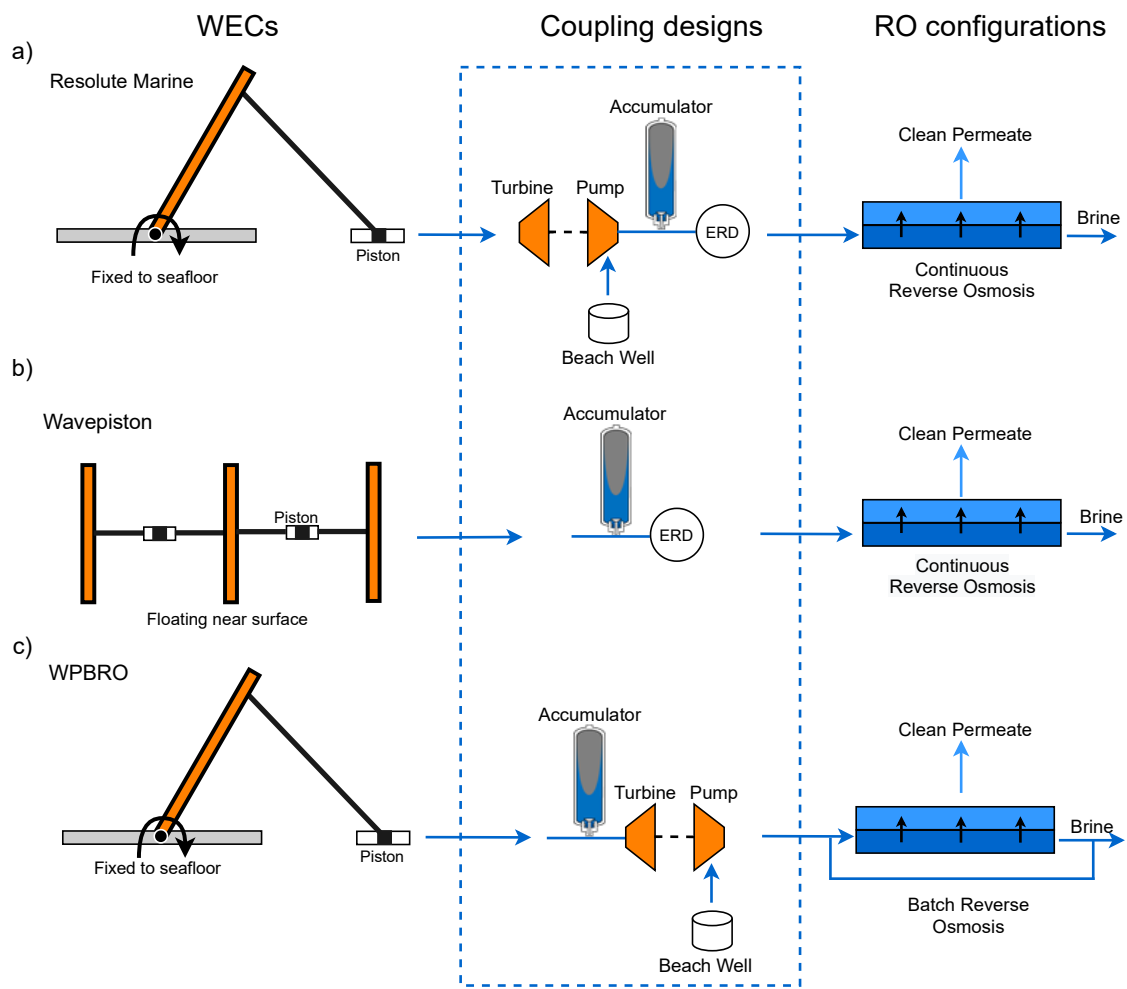

**Figure 1:** Comparison of (c) the WPBRO process with other state-of-the-art WEC-RO processes by (a) Resolute Marine [9] and (b) Wavepiston [34]. For each process, the sub-process configurations are displayed for the WEC, coupling, and RO system.

Modeling results predict that the SEC of the wave-powered batch reverse osmosis (WPBRO) system is 2.4 kWh/m$^3$ at the lowest predicted LCOW of $1.96/m$^3$ for a scale of 2400 m$^3$/day (Table 1). In contrast, energy estimates for leading configurations are shown in Table 1. Our WPBRO system shows promising second law efficiency, despite using more conservative/realistic assumptions for efficiency and driving pressure above osmotic (7.5) bar versus some past studies (1 bar, [37])

To be comparable with NREL's prior work [35], these results were determined for the sea state conditions representative of Humboldt Bay, California (Table S2). A sea state is defined in the model by wave height, peak wave period, and specification of either regular or irregular waves. Results were also determined for sea states in Greece and the British Virgin Islands, two potentially competitive markets for wave-powered desalination. By reducing energy consumption and complexity, WPBRO is promising for increased resiliency in coastal communities.

**Table 1:** The estimated values of the SEC from wave energy and second law efficiency for prior work in wave-powered reverse osmosis systems are presented in the table below. The least work for desalination was calculated based on the feed salinity and recovery, and then used to find the Second law efficiency. All temperatures were assumed to be 21.6 °C.

| System Type | Feed Salinity (g/kg) | Recovery Ratio | Assumed Component Efficiency | Predicted SEC (kWh/m³) | Second Law Efficiency | Reference |
|---|---|---|---|---|---|---|
| Surge WEC with pressure-exchanger energy intensifier | 35 | 0.45 | 85% | 2.8 | 37.7% | NREL [35] |
| Surge WEC with pressure-exchanger energy intensifier | 37 | 0.25 | 85% \| 90% | 2.1 | 45.6% | Folley et al. [37] |
| Wave-overtopping WEC with hydro-electric power generation | 35 | 0.45 | 73.5% | 2.9 | 37.0% | Contestabile et al. [31] |
| Surge WEC integrated with batch reverse osmosis at the same sea state as NREL | 35 | 0.5 | 85% \| 65% | 2.4 | 46.1% | WPBRO |

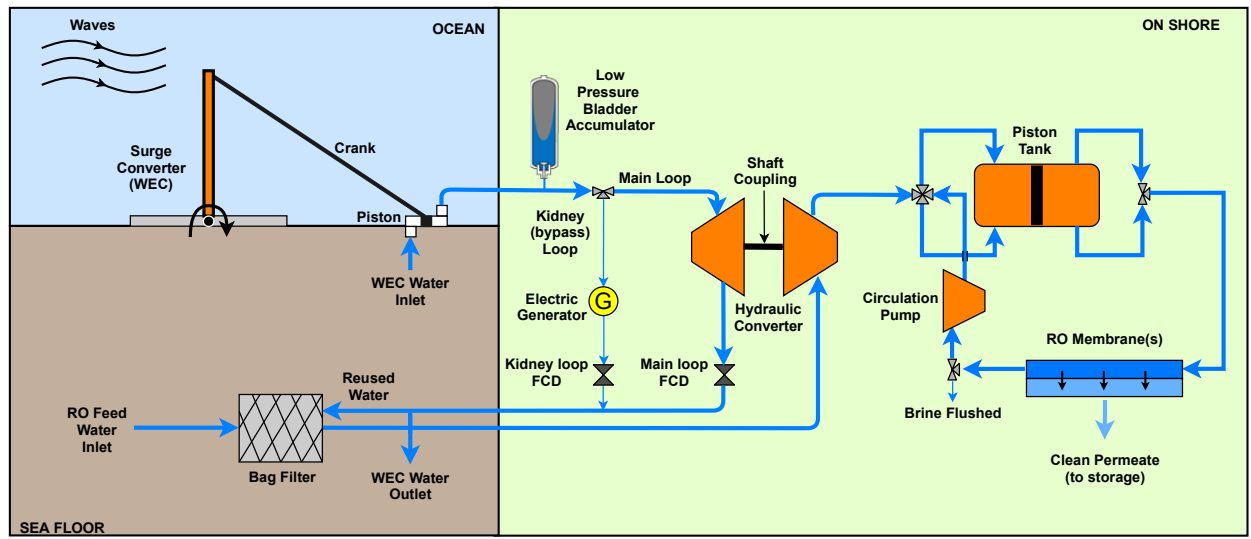

**Figure 2:** Simplified process diagram of novel WPBRO coupled via hydraulics. WPBRO pressurized water from the WEC (top left) is divided into a main loop (middle right) and a kidney (bypass) loop (far right). The main loop drives the turbine side of the hydraulic converter while the kidney (bypass) loop diverts flow from the main loop to an electric generator (center yellow) that powers the control system, the circulation pump in BRO, and a booster pump for the RO feed (not shown). RO feed water is drawn through a beach well (bottom left) and enters the pump side of the hydraulic converter (far right), pressurizing the water for BRO desalination. The direct use of this pressurized water to power BRO desalination eliminates any need for further energy conversion with pumps and motors (hydraulic - mechanical - electrical), thus reducing energy losses and increasing overall power available.

## Methods

The proposed WPBRO system is an integration of wave energy with BRO which includes a BRO system, a coupling (power take-off, PTO) system, and a WEC. This system was modeled and validated in MATLAB and Simulink, building off prior modeling of BRO [12] and of a wave energy to electric power system [38] created by Sandia National Laboratories and NREL. The model was developed through a series of governing equations and necessary assumptions and implemented as a time-domain simulation of wave-powered BRO. Optimization was done for different sea states which led to competitive SEC and LCOW results. The dynamic batch reverse osmosis model was validated with similar trends and values found in internally validated model results published by Park et al. (2021) [16]. The model is also comparable to Wei, et al. [39]. However, that model is designed on a smaller scale with lower feed salinity, a much lower maximum feed pressure, lower flow rates, and ideal pump efficiencies.

The system harnesses wave energy mechanically without a transition to and from electrical power, eliminating the need for an electrically driven high-pressure pump (the typical power generation device) in BRO. Instead, a hydraulic converter (Figure 4) is used to bring feedwater from atmospheric pressure to the required 30-70 bar for a BRO system following the osmotic curve. The slightly pressurized feedwater on the WEC-side provides energy to a turbine to directly drive the high-pressure pump side of the hydraulic converter for the BRO desalination process. Using a hydraulic converter allows for typical mid-pressure hydraulic wave devices to be compatible with a wide range of reverse osmosis pressures. A hydraulic accumulator dampens oscillations from the waves, and two flow control devices (FCD) are employed to maintain a constant accumulator pressure and a constant hydraulic converter shaft speed. In this model, the behavior of an FCD is like that of a throttling valve. The flow area through the component varies. A flow chart illustrating these components of the model is shown in Figure 3.

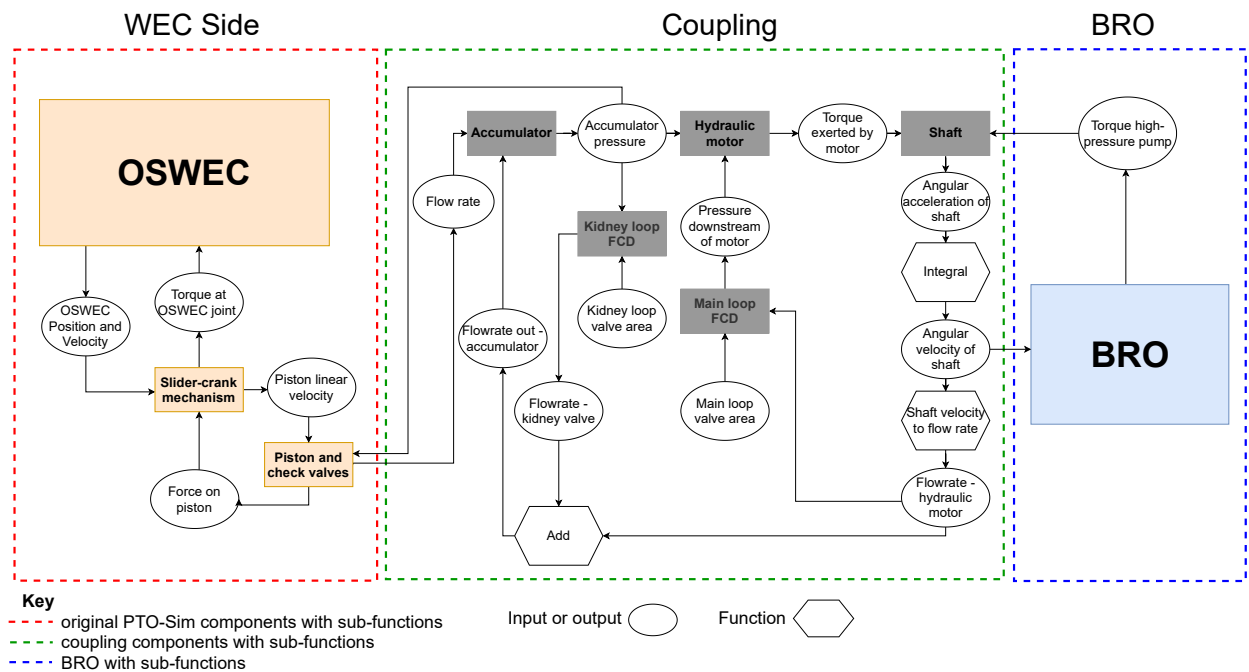

**Figure 3:** Wave-powered batch reverse osmosis Simulink model flow chart (excluding controllers). Each component in the system has interdependencies with other components. In the model, an oscillating surge wave energy converter (OSWEC) (red box, left) and associated power take-off components (slider-crank, piston) are connected to batch reverse osmosis and the high-pressure pump side of the hydraulic converter

(yellow box, right) via the proposed coupling (blue box, middle), which includes an accumulator to damp oscillations, a kidney (bypass) loop FCD to maintain the pressure in the accumulator, a turbine and shaft to represent the coupling side of the hydraulic converter, and a main loop FCD to maintain a constant shaft speed.

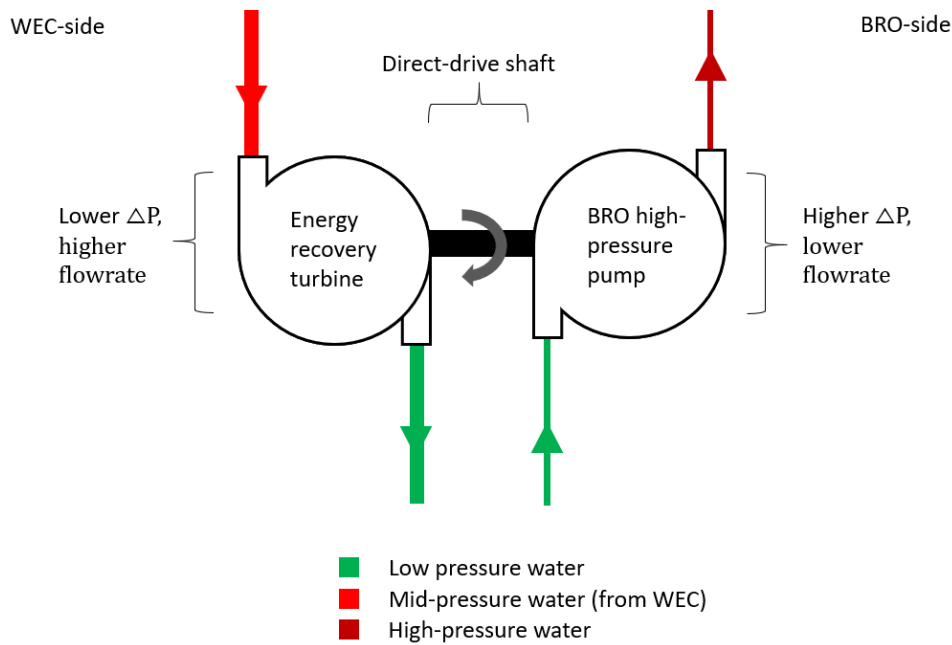

**Figure 4:** The hydraulic converter consists of a coupled turbine and pump for the WPBRO system. This device converts the hydraulic energy from the wave energy converter (high flowrate but not sufficient pressure for desalination), to be used at a higher pressure with a low flowrate.

The model builds on existing wave energy simulation tools (WEC-Sim and PTO-Sim) [38, 40]. WEC-Sim, and therefore the presented model, simulates the incoming waves using linear wave theory [40]. The WEC used in this model is the oscillating surge WEC (OSWEC) example from the WEC-Sim open-source repository (Figure 5) [41]. The OSWEC sits on the sea floor and acts like a flap, as shown in Figure 1. The WEC is connected to two pistons (Figure 3, piston and check valves, orange), which pressurize intake water as it is drawn from the seafloor, by a slider-crank mechanism [38]. By default, PTO-Sim tracks the performance of a system connecting a WEC with an electricity generation system [38], and it therefore provided a starting point for designing a WEC-to-desalination system. The referenced PTO-Sim model is a system that does not draw in seawater, instead using oil as a working fluid to be circulated through two accumulators [38]. Conversely, the proposed system opens the low-pressure sides of the pistons to draw in seawater, as done by Resolute Marine [9], which allows seawater to be the working fluid. Along with simplifying the system from a maintenance standpoint, this change is a more sustainable alternative to the closed-loop oil configuration.

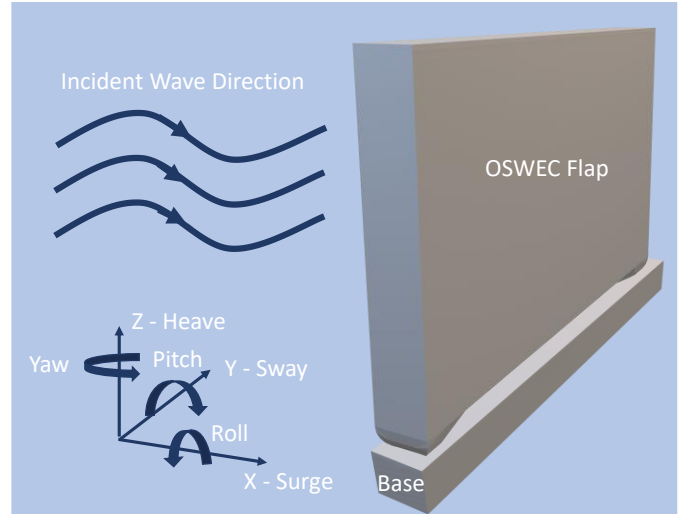

**Figure 5:** OSWEC geometry from WEC-Sim [41]. The base (lower rectangle) is secured to the seafloor, and the flap, upper rectangle swings back and forth generating energy from the waves.

**Governing Equations – WEC-BRO Coupling**

The equations below describe the modeling of the coupling system between the WEC and BRO. They are organized by referencing onshore components (Figure 2, green region), proceeding from left to right. These are followed by the equations governing BRO, specific energy consumption, and levelized cost of water.

*Accumulator*

The accumulator (Figure 2, top middle) dampens the highly oscillatory flow from the WEC. Similar to PTO-Sim, it is governed by the equation describing a polytropic process of an ideal gas (equation 1), where $n$ can be assumed to equal 1.4 for an adiabatic process [41]. The output flow from the accumulator is split between the main loop and the kidney loop.

$$V_{\text{accum}} = V_0 * \left(\frac{p_{\text{precharge}}}{p_{\text{accum}}}\right)^{\frac{1}{n}}$$
(1)

Here $V_{\text{accum}}$ is the instantaneous seawater volume in the accumulator (m³); $V_0$ is the initial seawater volume in the accumulator (m³) when the accumulator is empty of water; $p_{\text{precharge}}$ is the precharge pressure of the accumulator (Pa), the pressure of gas in the accumulator when it is empty of water as the initialization of the system is not modeled; $p_{\text{accum}}$ is the instantaneous pressure of the fluid in the accumulator (Pa), and $n$ is the adiabatic constant.

*Kidney (Bypass) Loop*

The purpose of the kidney loop FCD is to bleed off excess flow from the accumulator outlet, such that the accumulator remains charged below its maximum pressure and volume. Both FCDs are modeled using the orifice equation (equation 2) [42]. A turbulent-characteristic flow coefficient of 0.7 was chosen [42], and the density of seawater was assumed to be 1025 m³/kg. For this model, orifice size is synonymous with FCD area.

$$Q = C_{\text{f}} A_{\text{orifice}} \sqrt{\frac{2\Delta p_{\text{valve}}}{\rho}}$$
(2)

Here $Q$ is the flow rate through the FCD (m³/sec), $C_f$ is the flow coefficient, $A_{orifice}$ is the orifice size (m²), $\Delta p_{valve}$ is the pressure drop across the valve (Pa), and $\rho$ is fluid density (kg/m³).

*Power Transmission*

In this model, the hydraulic converter is modeled as a turbine connected to a high-pressure pump for BRO by a shaft (Figure 4). As water passes through the turbine in the main loop, hydraulic power is converted to mechanical power, and as water is drawn into the BRO-side, mechanical power is converted back to hydraulic power. The turbine is assumed to be a fixed positive displacement machine (equation 3) [42], and its shaft rotational velocity is governed by a torque balance (equation 4) like the shaft in [41], where back-torque from the high-pressure pump in BRO increases as the membrane pressure increases over a cycle [12].

$$NV_d = Q_{main} \tag{3}$$

Here $N$ is the shaft rotation rate (rev/s), $V_{d\_motor}$ is the volumetric displacement of the motor for one rotation of the shaft (m³/rev), and $Q_{main}$ is the flow rate through the main loop (m³/sec).

$$\frac{\tau_m + \tau_{hp}}{2\pi J} = \frac{\Delta p_{motor} V_d \eta_m/(2\pi) + \tau_{hp}}{2\pi J} = \frac{dN}{dt} \tag{4}$$

Here $\tau_m$ is the torque acting on the turbine (N-m), $\tau_{hp}$ is the torque acting on the high-pressure pump (N-m), $J$ is the rotational inertia of the shaft (kg-m²), $\Delta p_{motor}$ is the pressure drop across the turbine (Pa), $\eta_m$ is the motor efficiency, and $dN/dt$ is the shaft acceleration (rev/sec²) [42].

These equations influence the pressure on either side of the hydraulic converter as depicted in Figure 4 and thus affect the efficiency of the power take-off system. The pressure drop across the turbine side is directly proportional to the power transmitted by the turbine to the pump.

*Control of Kidney and Main Loop FCDs*

The basis for all FCD control is the orifice equation (equation 2). Because the turbine shaft rotation rate is directly proportional to its flow rate (equation 3), the orifice area is modulated to meet a desired turbine shaft rotation rate by way of proportional-derivative (PD) control [43]. First, the error between the desired and actual shaft speeds is computed (equation 5) [43]. Then, the control effort, change in FCD area, is found through the definition of PD control (equation 6) [43]. The controller gains were chosen experimentally to minimize settling time, overshoot, and chatter.

$$E(s) = R - Y = N_{shaft,ref} - N_{shaft} \tag{5}$$

Here $E(s)$ is the error, $R$ is the reference for control, $Y$ is the actual measurement of what is controlled, $N_{shaft,ref}$ is the desired shaft rotation rate (rev/s), and $N_{shaft}$ is the actual shaft rotation rate (rev/s).

$$\Delta A_{orifice} = E(s)C(s) = E(s)\big(K_p + K_d s\big) \tag{6}$$

Here $\Delta A_{orifice}$ is the necessary change in FCD area (m²), $E(s)$ is the control effort (m²-sec/rev, for the main loop FCD controller), $K_p$ is the proportional gain, $K_d$ is the derivative gain, and $s$ is the Laplace-domain variable for taking a derivative.

The kidney loop controller is also a PD controller, wherein the optimal controller gains were obtained experimentally. The error in the kidney loop is described by equation 7, and the change in area is described by equation 6.

$$E(s) = R - Y = p_{h,ref} - p_h \tag{7}$$

Here $E(s)$ is the error, $p_{h,ref}$ is the desired accumulator pressure (Pa) which is equal to the rated pressure of the accumulator, and $p_h$ is the actual accumulator pressure (Pa).

**Governing Equations – Batch Reverse Osmosis**

BRO is the most efficient RO desalination configuration realizable. The equations derived here build on previous work [12] and match the configuration in Figure 6.

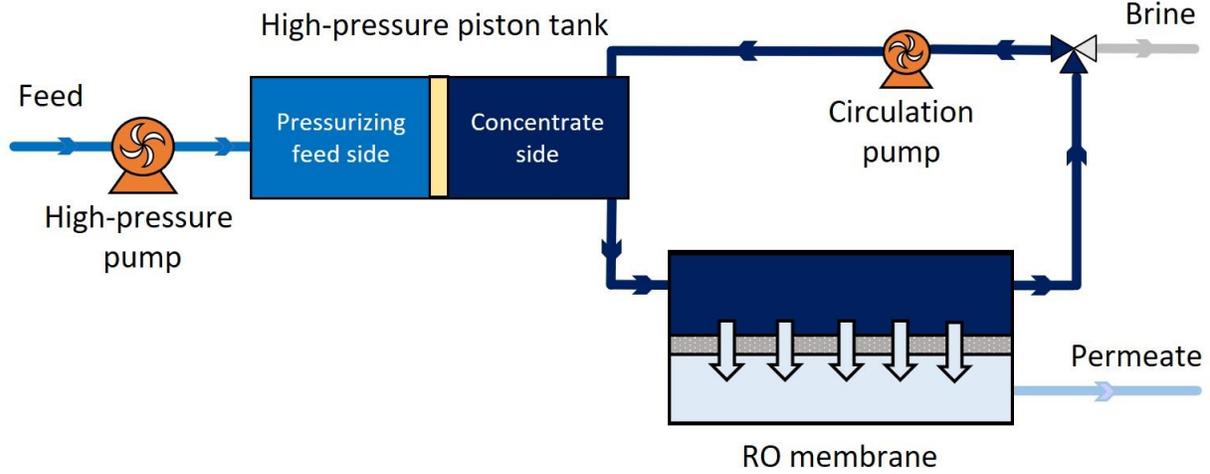

**Figure 6:** BRO featuring a double-acting high-pressure piston tank. Feed is brought up to the appropriate pressure via a high-pressure pump and used to push concentrate through RO membranes such that permeate is produced. This concentrate is then recirculated, so the salinity of the concentrate continuously increases throughout a cycle. This figure is presented in one of the author's prior work [12].

The high-pressure pump is modeled as a fixed-displacement machine (equation 8) and noting that the high-pressure pump flowrate is equal to the permeate flowrate by conservation of mass, the permeate flux is directly related to the shaft speed (equation 9) [42]. Equation 9 was developed for this model so that BRO could be scaled by changing the number of membrane modules in series and in parallel.

$$Q_{hp} = NV_{d\_pump} \qquad (8)$$

Here $Q_{hp}$ is the flowrate through the high-pressure pump (m³/sec), $N$ is the shaft angular speed (rev/s), and $V_{d\_pump}$ is the volumetric displacement of the high-pressure pump (m³/rev).

$$J_w = \frac{Q_p}{A_{mem}m_{ser}m_{par}} \qquad (9)$$

Here $J_w$ is the permeate flux through all membrane modules (m/s), $Q_p$ is the total permeate flow rate (m³/sec), $A_{mem}$ is the membrane area for one module (m²), $m_{ser}$ is the number of membrane modules in series, and $m_{par}$ is the number of membrane modules in parallel.

The osmotic pressure is the minimum membrane pressure required for reverse osmosis to occur. Permeate flux will occur for values of membrane pressure higher than the osmotic pressure (equation 10) [12]. Osmotic pressure increases throughout a BRO cycle as the membrane concentration increases. Note that the exponential term (equation 10) accounts for concentration polarization. The mass transfer coefficient was calculated using the Reynolds number. The Sherwood number correlation was obtained from [44].

$$\pi = iRT * C_{mem} * e^{\frac{J_w}{k}} \qquad (10)$$

Here $\pi$ is the osmotic pressure (Pa), $i$ is the van't Hoff factor, $R$ is the ideal gas constant (J/mol-K), $T$ is the fluid temperature (K), $C_{\text{mem}}$ is the bulk concentration of the fluid (g salt/kg water), and $k$ is the mass transfer coefficient (m/s).

The osmotic pressure is related to the feed-side pressure (equal to the high-pressure pump outlet pressure by a force balance), the permeate flux, and half of the pressure drop across all membrane modules in series (rightmost term of equation 11). This equation is derived from [44] and [45].

$$p_{\text{f}} = \frac{J_{\text{w}}}{\mathcal{A}_w} + \pi + \frac{f \rho u_{\text{avg}}^2}{4 D_{\text{h}}} L_{\text{mem}} m_{\text{ser}} \tag{11}$$

Here $p_{\text{f}}$ is the feed-side pressure (i.e. at the inlet of the membrane modules) (Pa), $\mathcal{A}_w$ is the membrane permeability (m/(s-Pa)), $f$ is the friction factor obtained from [45], $\rho$ is the fluid density (kg/m$^3$), $u_{\text{avg}}$ is the bulk fluid velocity across one membrane leaf (m/s), $D_{\text{h}}$ is the hydraulic diameter (m) equal to twice the spacer thickness, and $L_{\text{mem}}$ is the length of a single membrane module (m).

Instantaneous torque for a fixed-displacement pump is calculated last (equation 12), completing the connection between the coupling shaft speed and the torque on the BRO (high-pressure pump) side of the shaft [42]. As the load on the BRO system increases (i.e. osmotic pressure increases) at a constant shaft rotation rate, there is an increasing torque on the shaft connection between the high-pressure pump and the turbine. Here $\eta_{\text{hp}}$ is the high-pressure pump efficiency.

$$\tau_{\text{hp}} = \frac{V_{\text{d\_pump}} * p_{\text{f}}}{2\pi * \eta_{\text{hp}}} \tag{12}$$

Conservation of mass equations (water mass, salt mass) are used to update the volume and concentration of water in the piston tank over time. The volume and concentration are re-initialized at the end of each permeate production phase, allowing the simulation of multiple BRO cycles [44].

**Specific Energy Consumption**

The primary benefit of BRO is its ability to significantly reduce specific energy consumption (SEC), the energy consumed per unit mass of permeate produced (kWh/m$^3$) [12]. SEC was calculated in the model using equation 13, which encompasses all power types in the process.

$$SEC = \frac{avg(P_{\text{WEC}}) + avg(P_{\text{CP}})}{avg(Q_{\text{permeate}})} \tag{13}$$

Here $P_{\text{WEC}}$ is the power input from the WEC (W), $P_{\text{CP}}$ is the power input from the circulation pump (W), $Q_{\text{Permeate}}$ is the permeate output flowrate (m$^3$/s), and $avg(Q_{\text{permeate}})$ denotes the arithmetic average of a timeseries. Averages were computed to allow for simple comparison amid the oscillatory nature of the WEC power input and permeate output flow rate over time.

While initial modeling efforts focused on using two throttling valves as the FCDs, it was realized that both throttling valves incurred large power losses. To address this, a new system design was proposed in which both throttling valves would be replaced with electric generators as alternate FCDs with controllable counter electromotive forces. In principle, the generators would provide the same control effort as the throttling valves, but with the added benefit of reducing net power losses and consequently, SEC and the levelized cost of water (LCOW). While these controlled generator mechanisms were not developed in the WPBRO model, their effect was estimated by recovering the power losses in the valves and factoring in a generator efficiency [41]. The energy recovered by the generators is calculated with equation 14.

$$avg(P_{\text{recovered}}) = \eta_{\text{gen}}(avg(P_{\text{kidney loop valve}}) + avg(P_{\text{main loop valve}})) \tag{14}$$

Here $P_{\text{recovered}}$ is the power input from the generators (W), $P_{\text{kidney valve}}$ is the power lost in the kidney valve (W), and $P_{\text{main loop valve}}$ is the power lost in the main loop FCD (W). All power losses are calculated using the product of the flowrate through the component and the pressure drop across the component.

**Second Law Efficiency**

The second law efficiency is defined as a function of the least energy of separation and the specific energy consumption of the process. The least energy of separation is calculated via the isothermal Gibbs free energy description at a finite recovery.

$$w_{\text{least}} = g_{\text{p}} + \frac{1-r}{r} g_{\text{b}} - \frac{1}{r} g_{\text{f}}$$

Where $w_{\text{least}}$ is the least work in $\frac{\text{kJ}}{\text{kg}}$, $g$ is the specific Gibbs free energy [46], and r is the recovery ratio. The specific Gibbs free energy is a function of the salinity and temperature and is calculated by the MATLAB NaCl property libraries from [47]. The least work model with sodium chloride is compared to results with seawater with matching trends and values (S3). Subscripts signify the water stream where p is permeate, b is brine, and f is feed. The minimum specific energy consumption is found by a unit conversion of the least work.

$$SEC_{\text{least}} = \frac{\rho_{\text{water}} w_{\text{least}}}{3600}$$

Where $SEC_{\text{least}}$ represents the minimum specific energy consumption required for the process and $\rho_{\text{water}}$ is the density of permeate at the process temperature. The second law efficiency is defined as the ratio between the minimum energy required and the real process energy consumption.

$$\eta_{II} = \frac{SEC_{\text{least}}}{SEC_{\text{WPBRO}}}$$

Where $\eta_{II}$ is the second law efficiency. The second law efficiency is used to compare unlike processes on an equivalent basis.

**Economic Analysis**

The levelized cost of water (LCOW) of the WPBRO system with two generators was calculated using a method developed by NREL [35], clarified through a conversation with the authors. This method involves calculating the capital and operational expenditures of WPBRO and combining their effects through equation 15. In this work, all costs presented are in United States Dollars (USD).

*Capital Expenditures*

The financial analysis is comparable to NREL's paper which calculated the LCOW for a physical system corresponding to the original PTO-Sim model [35]. The WEC CapEx was assumed to be the same as NREL's WEC CapEx, $3,880,000 [35] because the WEC used in the WPBRO model was the same as the 18-meter wide WEC that NREL analyzed.

The CapEx for BRO can be estimated from a scaled budget of component parts (Table 2). For reference, an NREL study calculated an RO CapEx of $3,685,000 for a 3100 m³/day RO system [35]. For a 1700 m³/day RO system in Greece, the estimated CapEx for RO is $3,363,000 according to DesalData [48].

**Table 2:** Summary of BRO component costs for WPBRO system. Costs are based on quotes from various industry manufacturers and suppliers for a scale of 100 m³/day of permeate produced. The manufacturers

and part numbers for these quotes are referenced in Table S3. The shipping cost was determined by assuming shipment from West Lafayette, IN to the British Virgin Islands [49].

| System Component | Total Cost (USD) |
|---|---|
| Pumps | $25,000 |
| hydraulic converter (Energy Recovery Device) | $20,000 |
| Sensors | $10,000 |
| Valves | $20,000 |
| Pipes and Fittings | $12,000 |
| BRO Membranes (x8) | $6,000 |
| LP Bladder Accumulator | $33,000 |
| BRO Membrane Housing (x2) | $3,000 |
| Bag Filter | $3,000 |
| HP Piston Tank | $2,000 |
| Shipping | $12,000 |
| **Total Component Cost*** | **$146,000** |

*Before scaling to full 2400 m³/day capacity.*

NREL estimated the RO CapEx to be $3,685,000 for a 3100 $m^3$/day RO system [35]. Assuming a linear relationship with component cost, this amount divided by 31 yields an estimate of $118,861 for a 100 $m^3$/day RO system. The BRO CapEx for a capacity of 100 $m^3$/day was estimated as $146,000. Although slightly higher than the referenced work's estimate, the BRO cost includes a highly priced accumulator to account for sea water as the working fluid and to account for inflation. The BRO CapEx for 1700 – 2400 $m^3$/day capacity, the output of the model depending on sea state, could therefore be estimated in this work using linear extrapolation with a scaling factor determined by the amount of water produced. A study that assessed past RO desalination plant data to estimate CapEx showed a strong log-linear relationship between CapEx and plant capacity [50]. Using this model, the estimated BRO CapEx was confirmed.

*Operational Expenditures*

Operational expenditures for the WEC were also assumed to be the same as NREL assumed, $68,100 [35]. BRO system operational costs are dependent on permeate production capacity (m³/day). Table 3 lists how the operational costs of BRO are determined, where factors are identical to NREL's determination [35]. Labor costs are split between direct labor and management labor costs according to equations 7 and 8 in [35], where $Cap_{RO}$ is the capacity of 100 WPBRO systems in parallel. Annual water production (AWP) is calculated by multiplying the amount of water produced by 100 systems per day (m³/day) by the number of days in a year and a capacity factor, which accounts for the fact that the system has a significant amount of downtime.

**Table 3:** BRO OpEx Costs calculated from NREL's methodology [35]. OpEx is calculated for 100 systems in parallel, as is LCOW (each factor in the OpEx cost is scaled by a factor of 100).

| | |
|---|---|
| Direct Labor Costs | $29,700/laborer [35] |
| Management Labor Costs | $66,000/manager [35] |
| Spare Parts | $0.04/m³ * AWP * 100 |
| Pretreatment | $0.03/m³ * AWP * 100 |
| Posttreatment | $0.01/m³ * AWP * 100 |
| Membranes | $0.07/m³ * AWP * 100 |
| Insurance | 0.5% BRO CapEx * 100 |
| **Total** | **$161,000** |

Plant capacity is assumed to be 49% as found by NREL [35].

*Levelized Cost of Water*

The economic viability of WPBRO was measured by its LCOW, which estimates the overall cost for the system to deliver a cubic meter of water. The process of determining the LCOW of WPBRO was adapted from NREL's analysis of a WEC-RO system [35]. LCOW is found using equation 15.

$$LCOW = \frac{(FCR * CapEx) + OpEx}{AWP} \tag{15}$$

Here FCR is a fixed charge rate of 10.8% [49], CapEx is the total capital expenditure necessary to deploy the system, OpEx is the operational expenditures of the system per year, and AWP is the annual water production in m$^3$. A capacity factor of 49% was used for the system's production to account for changes in sea states, down times, and other losses not accounted for by mathematical modeling.

This final LCOW, $1.96 for a high energy sea state, is based on the late-stage re-design of replaced throttle valves with generators as the FCD devices. The LCOW of the competitive system in NREL's paper was found to be $1.82 [35]. The main factor that raises the WPBRO LCOW close to $2 is the high accumulator cost. The discrepancy of NREL's reported LCOW here compared to the LCOW in [35] was discussed with the authors. Further results for different sea states will be presented in the results section.

**Parameters**

The parameters used in the MATLAB and Simulink model are described below in Table 4 for reference.

**Table 4:** Modeling Parameters organized into groups based on their relevance to one another in MATLAB and Simulink. The references for each figure are listed in the last column; "design" indicates that the parameter was by design, custom for the WPBRO project. All unlisted parameters in the model are calculated in terms of the parameters listed here according to the equations above.

| Name | Variable | Value | Reference |
|---|---|---|---|
| OSWEC mass (kg) | n/a | 127000 | [41] |
| OSWEC height (m) | n/a | 8.9 | [41] |
| OSWEC width (m) | n/a | 18 | [41] |
| OSWEC thickness (m) | n/a | 1.8 | [41] |
| Slider-crank "crank" link length (m) | $R_2$ | 3 | [41] |
| Slider-crank "offset" distance (m) | $R_5$ | 1.3 | [41] |
| Slider-crank "rod" link length (m) | $R_3$ | 5 | [41] |
| Slider-crank piston area (m$^2$) | $A$ | .0378 | [41] |
| Accumulator initial volume (m$^3$) | $V_0$ | 6 | [41] |
| Accumulator rated volume (m$^3$) | $V_{rated}$ | 1.83 | [41] |
| Accumulator pre-charge pressure (Pa) | $p_0$ | 9.6e6 | [41] |
| Accumulator rated pressure (Pa) | $p_{rated}$ | 16e6 | [41] |
| Desired (reference) shaft speed (rev/s) | $N_{ref}$ | 50 | Design |
| Desired total recovery ratio (m$^3$ permeate/m$^3$ feed) | $RR_{tot}$ | .5 | [12] |
| Desired permeate flux (m/s) *corresponds to 30 LMH | $J_{w,des}$ | 8.33e-6 | [12] |
| High-pressure pump efficiency | $\eta_{hp}$ | .85 | [12] |
| Circulation pump efficiency | $\eta_{cp}$ | .65 | [12] |
| Turbine efficiency | $\eta_m$ | .95 | [41] |
| Hydraulic generator efficiency | $\eta_{gen}$ | .85 | [41] |
| Membrane brine spacer thickness (m) | $\delta$ | 7.112e-4 | [12] |
| Membrane length (m) *for one module | $L_{mem}$ | .96 | [12] |

| Membrane area (m²) *for one module | $A_{mem}$ | 7.4 | [12] |
|---|---|---|---|
| Desired recovery ratio per pass (instantaneous) (m³/s permeate / m³/s feed) | $RR_{inst,des}$ | .1 | [12] |
| Desired proportion of total flowrate through main loop *reduced from .9 due to additional factor of safety | $p_{main}$ | .8 | Design |
| VDC tank length (m) *volume and area vary to meet the desired total recovery ratio | $L_{tank}$ | 1 | [12] |
| Membrane water permeability coefficient (m/s/Pa) | $A_w$ | 5.56e-12 | [12] |
| Kinematic viscosity of water (m²/s) | $\nu$ | 8.56e-7 | [12] |
| Diffusion coefficient of salt in water (m/s) | $D$ | 1.47e-9 | [12] |
| Initial feed salinity (g salt / kg water) | $C_{f,i}$ | 35 | [12] |
| Molar mass of NaCl (kg/kmol) | $M_{NaCl}$ | 58.55 | [12] |
| Feedwater density (kg/m³) | $\rho$ | 1025 | [12] |
| Universal gas constant (kJ/kmol/K) | $R$ | 8.314 | [12] |
| System temperature (K) | $T$ | 300 | [12] |
| van't Hoff factor for NaCl | $i$ | 2 | [12] |
| Osmotic coefficient of membrane | $\Phi$ | .93 | [12] |
| Shaft rotational moment of inertia (kg*m²) | $J$ | 2000 | Design |
| Main loop valve proportional gain (m²/(rev/s)) | $K_{p,m}$ | 1e-5 | Design |
| Main loop valve derivative gain | $K_{d,m}$ | 100e-5 | Design |
| Kidney valve proportional gain (m²/Pa) | $K_{p,k}$ | -6.67e-14 | Design |
| Kidney valve derivative gain | $K_{d,k}$ | -667e-14 | Design |
| Valve flow coefficient | $C_f$ | .7 | [42] |

The desired reference shaft speed was chosen to correspond with a reasonable physical shaft speed for rotary machines [52]. The desired proportion of flow through the main loop was chosen to ensure that there would always be positive flow through the kidney loop, accounting for fluctuations in the input flowrate. Originally, a proportion of 0.9 was chosen, but a proportion of 0.8 led to a more stable system design. To avoid discontinuities in the required control effort, the shaft inertia needed to be high, so it was increased to 100 times the shaft inertia in PTO-Sim [41]. As will be mentioned in Assumptions and Constraints, in future work, the mass and geometry of physical components should be referenced. The final four design parameters, the controller gains, were iterated until finding the optimal values to minimize settling time, overshoot, and chatter, as mentioned in Methods: Control of Kidney and Main Loop Valves.

**Assumptions and Constraints**

On the WEC-side of the model, the following assumptions were made. In accordance with WEC-Sim, the WEC component was modeled using linear-wave theory, including added mass, radiation damping, and wave excitation forces [40]; further, irregular waves were modeled as a superposition of regular waves using a spectrum of characteristic frequencies [40]. All feedwater coursing through the system was assumed to be incompressible and flow was assumed one-dimensional and uniform. Losses within the pipes were considered negligible, and the pipes were assumed to have negligible volume. This assumption was deemed reasonable due to the high flow rates on both sides of the system. The high flow rates correspond to a high Reynolds number, which is inversely proportional to friction factor, the latter of which is directly proportional to head loss in the piping according to the Darcy-Weisbach formula. The head loss is therefore expected to be very small [53]. Gas within the accumulator was assumed to be an ideal gas [42]. Prior to running the model, it was assumed that the WPBRO system was pre-charged to desired initial conditions with the desired shaft speed, accumulator rated pressure and volume, and both valve areas initialized. All pumps and motors were assumed to be fixed displacement machines with no volumetric losses, and the sea state input to WEC-Sim was assumed to persist for 24 hours. Furthermore, the control valves (Figure 1, throttling valves) were assumed to draw negligible power input.

The BRO-side shares all assumptions with the WEC-side in addition to the following: The flushing step was assumed to have negligible duration. This assumption is justified as flushing can be assumed to occur at 10 times the normal flowrate [12]. The model calculated that compared to the cycle time, this interval was small. However, flushing takes a nonzero amount of time to occur, and additional system considerations would be necessary to implement flushing in this coupled WPBRO system, especially due to mixing effects [39]. Accounting for flushing would provide a more realistic estimate of how the system control effort would change over time. The flushing step, or lack thereof, is currently the most uncertain piece of the model and should be implemented and studied in depth in future iterations of the model. The current model only considered flushing minimally to focus the scope on the active stage of BRO.

Additionally, on the BRO-side, mixing in the high-pressure pump was assumed to be instantaneous, such that the bulk concentration on the active side of the tank is uniform. Acceleration of the piston in the tank was also assumed to be zero. For membrane configurations, it was assumed that flow is identical for all branches in parallel, and it was assumed that bulk concentration increases linearly as flow progresses through a branch. Bulk parameters were approximated as the average of conditions at the inlet and outlet of the branch. Cycle-to-cycle salt retention was not considered in this model, although recent experimental work showed that this may be near 5% depending on pipe size and feed salinity [39].

A few constraints of the model should be considered in future work in addition to flushing, including reconfiguration to predict instead of manually determine the ideal number of membrane modules and referencing specific physical components for some model parameters.

Due to the design of the model, the following must be true for every time step in the simulation. First, the flow rate through the kidney loop must be greater than zero for all time – if the flow rate though this loop is ever negative, it corresponds to a condition where flow is flowing from the kidney loop outlet into the main loop, which is nonphysical. This result can occur when waves are highly irregular. Second, the pressure at the main loop FCD inlet must be greater than zero for all time. If this pressure goes to zero, the main loop FCD will be unable to exert control effort. For example, in the valve FCD system, if the pressure drop is near zero, the valve area will approach infinity, which is nonphysical. This result can occur when the load on the BRO side (Figure 2, right of hydraulic converter) is too large for a given sea state. One factor that increases BRO load is an increase in the number of membrane modules – hence, the main loop FCD constraint provides an upper limit to the number of modules that can be added in the model.

A more robust system would consider the variation in wave conditions to automatically, as opposed to manually, determine the optimal configuration of membrane modules while minimizing the specific energy consumption. The waves were modeled as irregular by nature to more accurately replicate reality. This made it difficult to predict the number of membranes for precluding a zero-pressure scenario at the main loop FCD.

In addition, in calculating motor displacement volume and selecting a shaft mass, some sizing parameters of real-world components were not directly referenced – valve areas, shaft size and mass. Future work should impose more rigorous constraints on sizing by selecting dimensions based on off-the-shelf products.

## Results

The MATLAB and Simulink model for WPBRO indicated similar physical trends to PTO-Sim [54] and modeling of BRO. Flow power through different components and the building of pressure over time in BRO were especially significant findings. Furthermore, implementing wave-powered BRO with generators instead of throttling valves (WPBRO-Gen), yielded lower SEC and LCOW values. The generators increased the power take-off efficiency of the coupling. Notably, recovery ratio per pass on the BRO side also influences SEC. The model was found to handle a range of different sea states, numbers of membranes in parallel, and permeate fluxes, and the controllers tested proved functional.

### Validation

Both the wave (WEC) and desalination (BRO) results were validated against independently published modeling dynamic models, as well as specific energy consumption results from previous experiments. To validate the model on the WEC-side, its outputs were compared to existing PTO-Sim modeling work [54], which itself has been experimentally validated. The WEC flow power graph of the previously published results and this model appear similar in magnitude and frequency (Figure 7a). In addition, the component flow powers verify that energy is conserved on the WEC-side: the WEC power is equal to the sum of motor power, kidney valve power, and main loop FCD power. On the BRO-side, membrane feed and osmotic pressures were plotted over time and the resulting graph bears similarity to the pressurization behavior of the BRO process [12] (Figure 7b) [55].

The next step in validation was ensuring that the controllers functioned as desired. The main loop FCD control effort was evaluated based on how well the hydraulic converter shaft speed adhered to its set point. As the kidney (bypass) FCD controlled the pressure within the accumulator, its control effort correlated to how well-dampened the pressure was and how accurately the accumulator pressure was kept around the rated pressure. Both control efforts are functional as the main loop FCD drove shaft speed to a constant value (Figure 7c) and the bypass valve held the accumulator pressure around 16 MPa (Figure 7d).

Our model predicts that the batch reverse osmosis portion of the model achieved an SEC of $2 - 2.2$ kWh/m^3. This estimate is predicted to be slightly above the 1.7-1.9 kWh/m^3 range that was shown by Wei, et al (2021) [39]. With both model and approximate experimental validation, the batch reverse osmosis model shown in this paper may serve as a conservative estimate of energy consumption and efficiency.

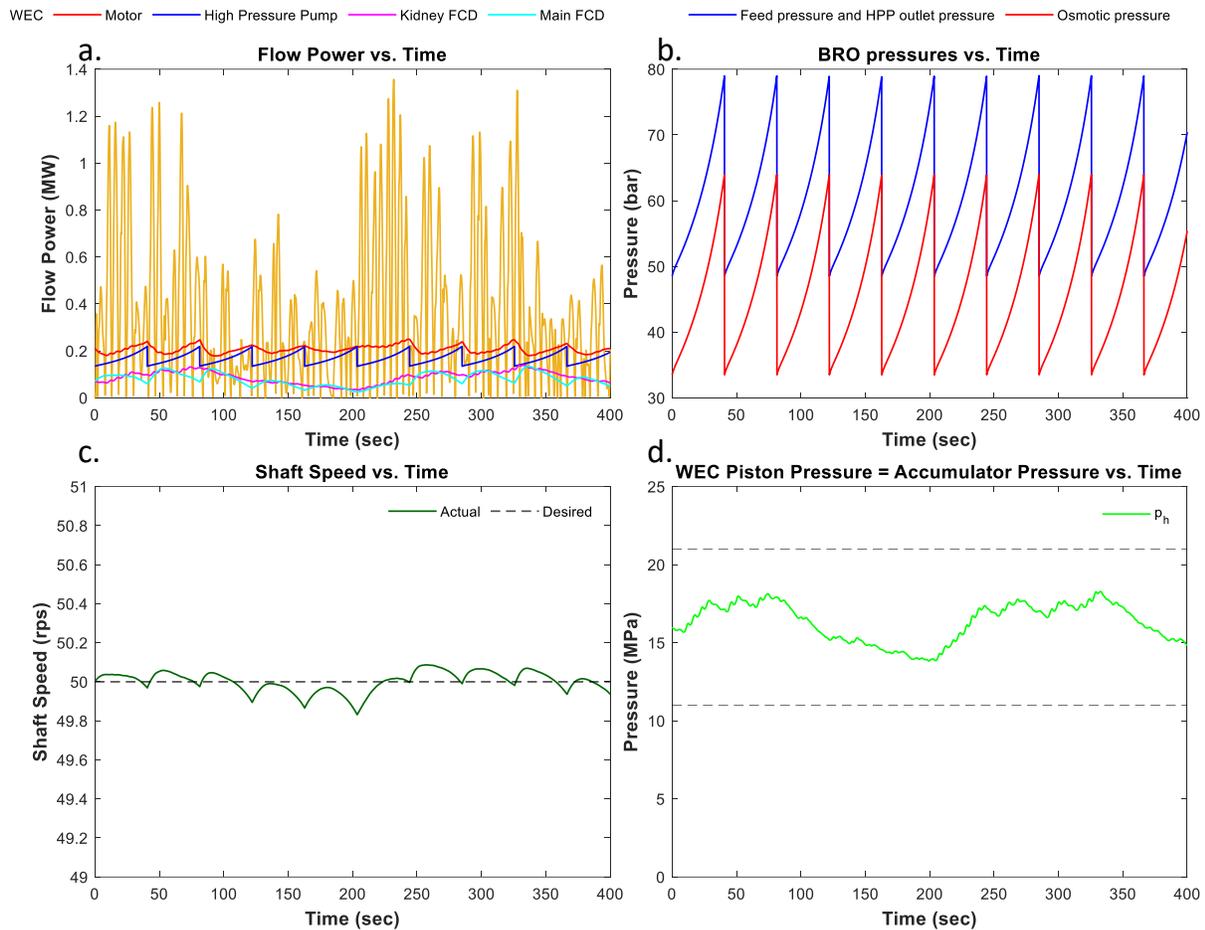

**Figure 7:** Key power conversion and pressure subsystems operating vs. time. All modeled with wave power under an irregular sea state condition with a wave height of 3 meters and a peak period of 11 seconds. (a) WEC-side energy is conserved, (b) BRO-Side pressure builds over time, (c) Main loop FCD controls shaft speed, (d) Kidney loop FCD controls accumulator pressure. These plots validate the model as explained according to PTO-Sim results [54], BRO expectations [12], and control theory [43].

If the FCDs double as throttling valves or hydroelectric generators, the SEC is competitively low at only 2.3 kWh/m³ for the wave-powered BRO with generators (WPBRO) system, seen in Figure 8. In this figure, a comparison to other systems can be seen as well, namely a WEC-RO system and an electricity to RO (Elec-RO) system [8]. When incorporating energy reuse in the WPBRO configuration, the system is much more efficient than these prior systems [8].

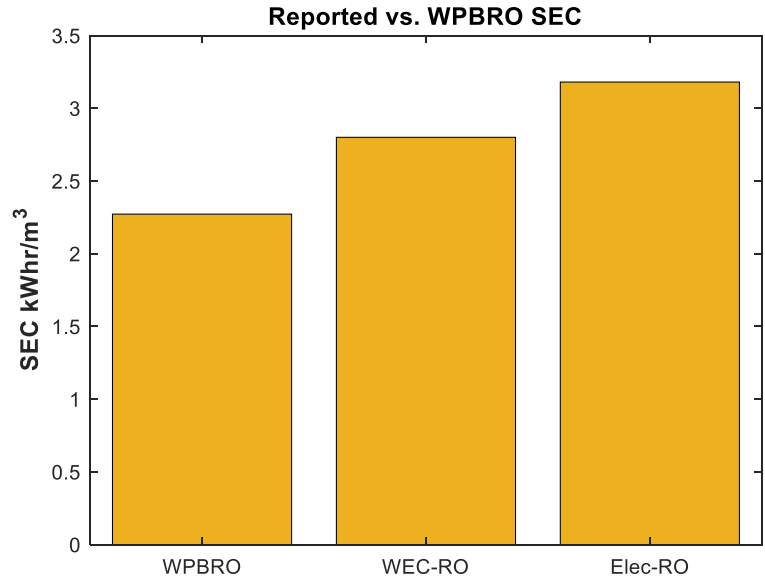

**Figure 8:** Specific energy consumption (SEC) of the proposed WPBRO system vs. main competitors. Includes proposed system with the generators replacing valves (WPBRO, far left), a WEC-RO system (center), and a wave-powered electricity, "Elec-RO" system (right) [8]. The WPBRO value was calculated at an irregular sea state of wave height of 3 meters and a wave period of 11 seconds.

Throughout testing, a tradeoff between SEC and permeate flux was observed: as permeate flux increases, SEC increases. Setting the permeate flux to a higher value leads to more energy intensive processes on the BRO-side and a higher membrane rated pressure is required. The contour plots of WEC efficiency versus recovery ratio per pass for a given setpoint flux illustrate this trend (Figure 9), where SEC is proportional to wave energy requirement. An instantaneous recovery ratio per pass of 0.1 m³/s permeate / m³/s feed (Table 3) was used in the model. To optimize recovery ratio per pass, it could be calculated as a function of flux instead of explicitly defined. The most optimal recovery ratio per pass would be obtained through a combination of membrane modules in series and in parallel since flux is a function of these parameters.

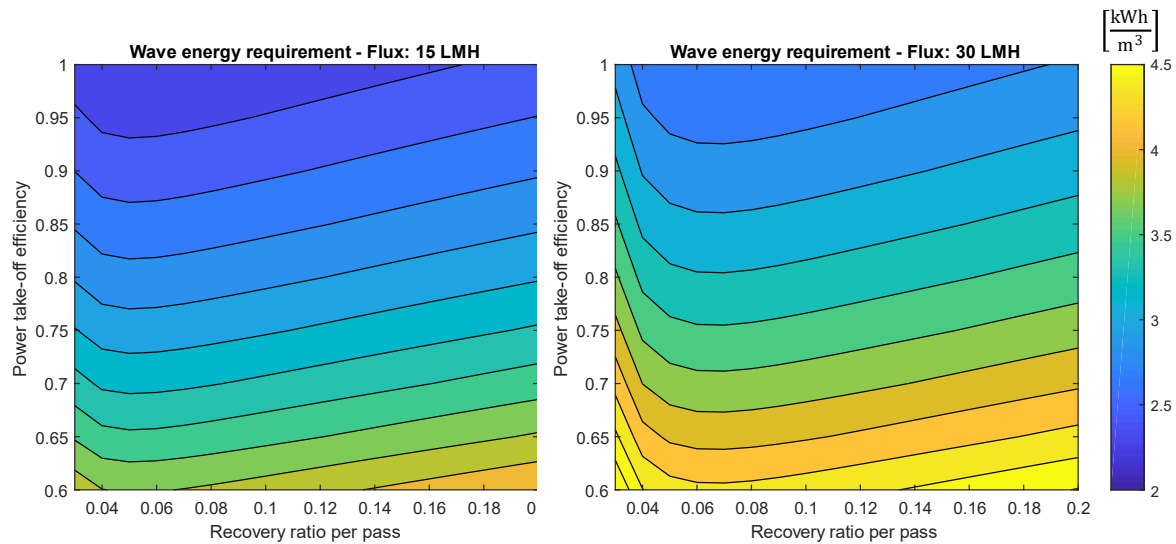

**Figure 9:** Wave energy required based on permeate flux, power take-off efficiency, and recovery ratio

per pass for a given average membrane water flux. For a smaller flux (left) and for greater power take-off efficiencies, SEC drops.

**Sea State Variation**

The model is robust and handles a multitude of WEC and RO inputs. WPBRO was tested with five representative sea states [54, 55], and the outputs correlate well with the expectation that higher energy sea states will result in higher permeate production, higher SEC, and lower LCOW.

The most energy-dense sea state in Humboldt Bay, California, was chosen as a benchmark to compare to existing PTO-Sim literature [54]. Sea states in Kos, Greece, and Guana Island, BVI [55] were chosen to see how less energy-dense sea states would affect the model. Table S2 presents the tested sea states.

The results for each sea state are shown in Table 6, and a bar chart showing how SEC and LCOW vary with sea state and the control type (valves vs. generators) is also telling (Figure 10). The number of membrane modules was configured for each sea state to produce the most water possible while maintaining the kidney loop flowrate above zero and the absolute pressure of the fluid in the main loop above zero. The number of modules was varied in intervals of 10 until the minimum motor valve power loss was significantly above zero. SEC clearly increases with greater wave energy density, and LCOW has the opposite effect, decreasing with greater wave energy density. Thus, as wave energy density increases, the WPBRO system is more cost effective but does not use energy as efficiently.

**Table 6:** Osmo-Sim results for the various sea states in table 4. A bar graph depicting the SECs and LCOWs for each sea state, both with the valve model and with the generator model, is shown below in Figure 10. Sea state is given in terms of wave height (meters) and peak wave period (sec).

| Sea State | Units | 3 meters, 11 sec | 1.5 meters, 6.75 sec | 1 meter, 5.5 sec | 1.75 meters, 9.25 sec | 1.25 meters, 7.25 sec | Reference |
|---|---|---|---|---|---|---|---|
| Energy density of sea state (kW/m) | kW/m | 48600 | 7500 | 2700 | 13900 | 5600 | [54] |
| Number of membrane modules in parallel | -- | 450 | 320 | 320 | 320 | 320 | Design |
| Average WEC input flowrate | m$^3$/s* | .022 | .014 | .014 | .014 | .014 | Result |
| Hydraulic main displacement volume | m$^3$/rev | 3.52e-4 | 2.24e-4 | 2.24e-4 | 2.24e-4 | 2.24e-4 | Design |
| High-pressure pump displacement volume | m$^3$/rev | 5.55e-4 | 3.95e-4 | 3.95e-4 | 3.95e-4 | 3.95e-4 | Design |
| VDC volume | m$^3$ | 1.18 | .84 | .84 | .84 | .84 | Design |
| Minimum power lost in main loop FCD as valve | kW | 25±20 | 14±4 | 14±4 | 14±4 | 14±4 | Result |

| Average output permeate flowrate | m³/day | 2400 | 1700 | 1700 | 1700 | 1700 | Result |
|---|---|---|---|---|---|---|---|

*Varies intensely for 3 meters, 11 sec, ±.006.

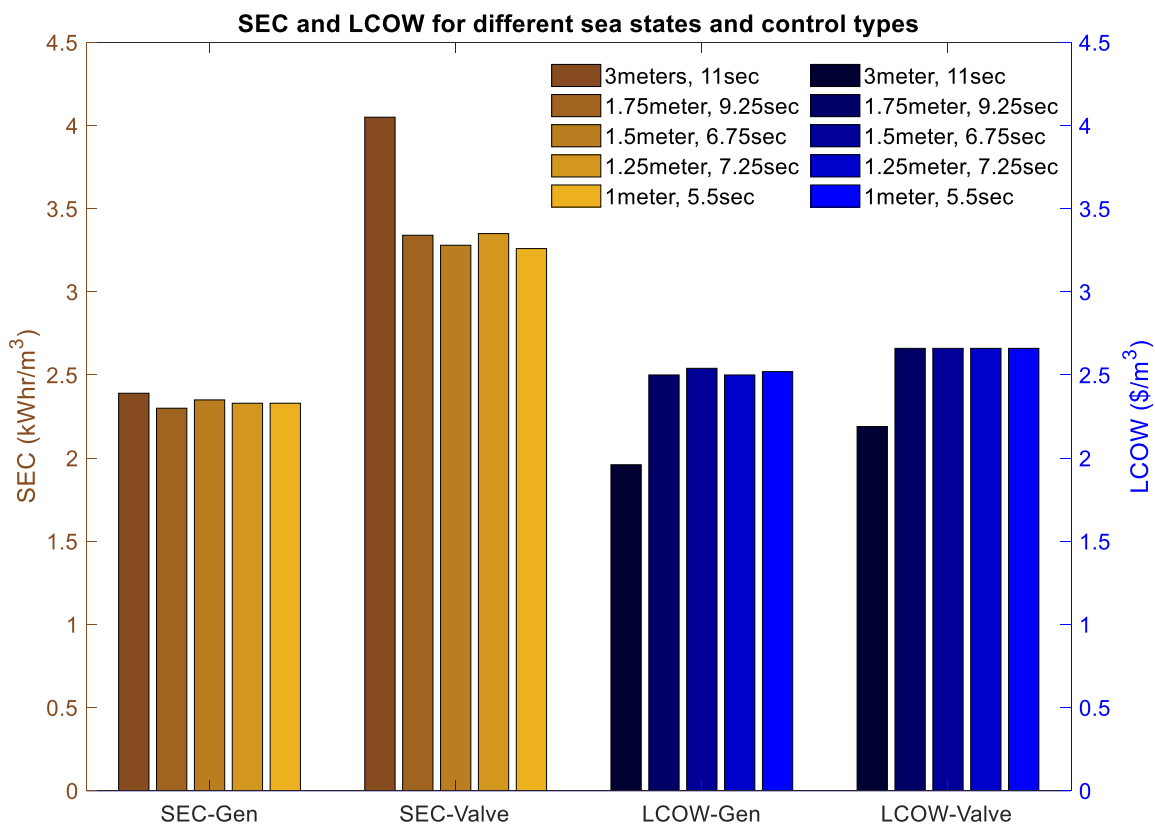

**Figure 10:** Sea state bar graph for the sea states in Table 4, such that the results in the last four rows of Table 5 can be visualized. SEC and LCOW are both higher with valve control, as opposed to generator (gen) control, and while SEC increases with more energy dense sea states, LCOW decreases. Recall, NREL predicts an LCOW of $1.82 for a wave-powered industrial RO system [35]. Exact values are presented in Table S1.

## Conclusions

This work analyzes the first WPBRO system, which includes a novel hydraulic converter to couple wave energy with BRO without electricity generation and uses seawater as an environmentally friendly working fluid. Dynamically coupling wave energy with BRO can lead to competitive system designs, compared to existing methods. The model of the proposed technology is robust and demonstrates that a WPBRO system can handle various sea states to produce 1700-2400 m³/day with an SEC ranging from 2.30-2.39 kWh/m³, with generators as FCDs. The predicted LCOW for this system ranges from 2.34-1.96 $/m³ in this generator configuration. When compared at the same sea state, the predicted levelized cost of the proposed WPBRO system ($1.96/m³) is competitive with previous state-of-the-art wave-powered desalination ($1.82/m³) [35]. Improving the economics of wave energy converters will significantly improve the cost effectiveness of wave-powered desalination as well. Future work on WPBRO should include detailed modeling of the flushing stage, transport mechanisms, system optimization for number of membranes and component

parameters, and control with generator electromotive force, as well as further study of pre-treatment in low-salinity and down-time in the BRO process.

## Acknowledgements

The authors would like to thank Abhimanyu Das, Antonio Esquivel Puentes, and Sandra P. Cordoba for their assistance with hydraulic modeling and Dr. Matt Folley for his assistance with hydrodynamic modeling. The authors are grateful for the DOE and NREL's Marine Energy Collegiate Competition for providing the structure and funding (SUB-2021-10582) that guided this work, and for awarding the Purdue team first place in their Marine Energy Collegiate Competition. The authors would like to thank the Bureau of Reclamation (R21AC10168-00), Purdue EVPRP, and Bob Boulware for funding this work, as well. A.R. would like to thank the Purdue Mechanical Engineering department.

## Conflict of Interest

The authors have multiple patents on batch reverse osmosis technologies, and Author Katherine Brodersen works for a wave-powered desalination company, Oneka Technologies

## Supplemental

The supplemental materials include a link to GitHub for downloading the MATLAB and Simulink model, detailed SEC and LCOW numbers from Figure 10, the details on the sea states tested, and the breakdown of BRO component costs. The manufacturers and part numbers are included for validation and future reference.

The MATLAB and Simulink model can be found at https://github.com/PurdueMECC/MECC2021_code

### S1. Specific energy consumption and levelized cost of water data

**Table S1: Values for SEC and LCOW values as presented in Figures 7 and 9 above. This table is an extension of Table 5.**

| Sea State | 3 meters, 11 sec | 1.5 meters, 6.75 sec | 1 meter, 5.5 sec | 1.75 meters, 9.25 sec | 1.25 meters, 7.25 sec |
|---|---|---|---|---|---|
| SEC with gen (kWh/m$^3$) | 2.39±.02 | 2.35±.01 | 2.33±.01 | 2.30±.01 | 2.33±.01 |
| SEC no gen (kWh/m$^3$) | 4.05±.12 | 3.28±.05 | 3.26±.02 | 3.34±.08 | 3.35±.03 |
| LCOW with generator FCD ($/m$^3$) | $1.96 | $2.54 | $2.52 | $2.50 | $2.50 |
| LCOW with valve FCD ($/m$^3$) | $2.19 | $2.66 | $2.66 | $ 2.66 | $2.66 |

### S2. Sea state variation

**Table S2:** Sea state conditions for Kos, Greece, Guana Island, British Virgin Islands (BVI), and Humboldt Bay, California [54, 55]. The significant wave height (H$_s$) and peak period (P$_p$) is listed for each location in boreal winter (December-January-February) and boreal summer (June-July-August). For each sea state, the wave energy density (J) was calculated [56]. The most energy-dense location was Humboldt Bay, California.

| | Boreal Winter | Boreal Summer |
|---|---|---|
| **Kos, Greece** | | |
| $H_s$ | 1.5 m | 1 m |
| $P_p$ | 6.75 s | 5.5 s |
| $J$ | 7,451 kW/m | 2,698 kW/m |
| **Guana Island, BVI** | | |
| $H_s$ | 1.75 m | 1.25 m |
| $P_p$ | 9.25 s | 7.25 s |
| $J$ | 13,898 kW/m | 5,558 kW/m |
| **Humboldt Bay, CA** | | |
| $H_s$ | 3 m | |
| $P_p$ | 11 s | |
| $J$ | 48,570 kW/m | |

**Table S3:** Summary of BRO component costs for WPBRO system. Costs are based on quotes from various industry manufacturers and suppliers for a scale of 100 m³/day of permeate produced. The manufacturers and part numbers for each component are referenced.

| System Component | Total Cost (USD) | Manufacturer | Part Number |
|---|---|---|---|
| Centrifugal Booster Pumps | $3,000 | Dayton Water Systems, West Carrollton, Ohio | 2ZWX1A |
| Axial Piston Circulation Pump | $22,000 | R.S. Corcoran Company, New Lenox, IL | 2000 F-HD3 |
| Hydraulic converter (Energy Recovery Device) | $20,000 | Energy Recovery, San Leandro, CA | 15000-20000 |
| Pressure and Flow Sensors | $7,400 | Keyence, Ithaca, IL | GP-M100 (Pressure), FD-Q50C (BRO Flow), FD-R80 (WEC Flow), OP-85502 (Pressure Cable), OP-87274 (Flow Cable) |
| pH Sensors | $1,300 | Sensorex Corporation, Garden Grove, CA | SD7000CD |
| Temperature Sensors | $300 | Atlas Scientific LLC, New York, NY | ENV-50-TMP |
| Conductivity Sensors | $800 | Omega Engineering Inc., Norwalk, CT | CDCE-90-001 |
| Check Valves | $3,500 | Sharpe, Northlake, IL | SMCSV25116024 |
| Throttling Valves | $5,500 | DynaQuip, St. Clair, MO | E3S2AAJE02 |
| Pressure Relief Valves | $7,500 | McMaster-Carr, Elmhurst, IL | 4256T16 |
| Ball Valves | $3,500 | Ashured Automation, Roselle, NJ | G33DAXR4F |
| Pipes and Fittings | $12,000 | Titan Fittings, Denver, CO | N6S8-16-03 (Nipple), SS-5605-16-16-16 (Tee), SS-9037-12-12 (Adapter), Swivels: SS-1502-16-16, SS-5000-16-16, Pipes: SS-5406-16-04, SS-5405-08-16, SS-5406-16-08, SS-5406-16-12, SS-5404-16-16, SS-1404-16-16, SS-5406-20-16, SS-5404-20-16 |
| | | W. W. Grainger, Inc., Lake Forest, IL | 1MKJ6 (Tank Fitting), 1RTW4 (Bushing), 2TV85 (Bushing), Pipes: 1LUP6, 2TY85, 2TV85 |
| Manifolds | $200 | Pneumadyne Inc., Plymouth, MN | 2KHL7 |
| BRO Membranes (x8) | $6,000 | FilmTec, DuPont Water Solutions, Santa Clara, CA | SW30HRLE-440 |
| BRO Membrane Housing (x2) | $3,000 | Pentair, Minneapolis, MN | GA90145-31 |
| LP Bladder Accumulator | $33,000 | IFP Automation, Brownsburg, IN | 109.622-02020 |
| Bag Filter | $3,000 | Forsta Filters Inc., Los Angeles, CA | M2-90 316L |

| HP Piston Tank | $2,000 | JIT Cylinders Inc., Hartselle, AL | 20002136 H-MX0 |
|---|---|---|---|
| **Total Component Cost\*** | **$146,000** | | |

*\*Before scaling to full 2400 m³/day capacity.*

## S4. Least work validation

Calculating the 2$^{nd}$ law efficiency is dependent on the process agnostic calculation of least work. In this work, we use a MATLAB implementation of the Pitzer model for aqueous NaCl mixtures to find the Gibbs free energy [47]. The least work using the NaCl properties is compared below with results using seawater properties from Mistry, et al. (2013) [46].

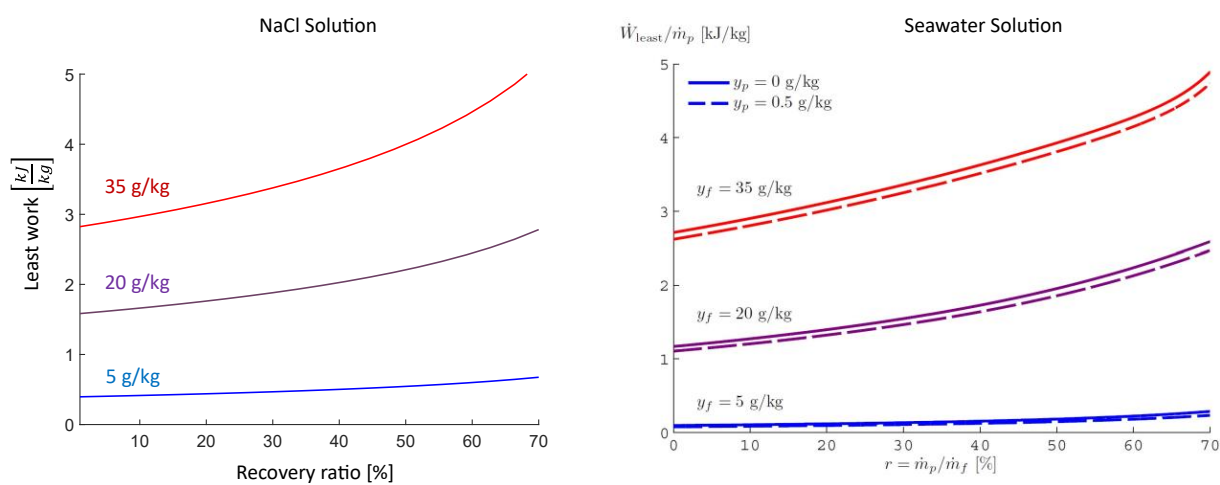

**Figure S4:** The NaCl solution (left) is computed via a Pitzer formulation. The seawater solution (right) is figure 3, directly from Mistry et al (2013). Numerical values and trends closely match between the two implementations.